\documentclass[aps,prl,twocolumn,amsmath,amssymb,footinbib,superscriptaddress]{revtex4}
\usepackage{graphicx}
\usepackage{epstopdf}
\usepackage{hyperref}
\usepackage{lpic}

\newcommand{\fig}[1]{Fig.~\ref{#1}}

\usepackage{color}
\newcommand{\tms}[1]{#1}

\begin{document}

\title{Foliated Quantum Codes}

\author{A. Bolt}
\affiliation{ARC Centre for Engineered Quantum System, Department of Physics, University of Queensland, Brisbane, QLD 4072, Australia}
\author{G.  Duclos-Cianci}
\author{D. Poulin}
\affiliation{D\'epartement de Physique, Universit\'e de Sherbrooke, Qu\'ebec, Canada}
\author{T. M. Stace} \email[]{stace@physics.uq.edu.au}
\affiliation{ARC Centre for Engineered Quantum System, Department of Physics, University of Queensland, Brisbane, QLD 4072, Australia}
\date{\today}
\frenchspacing

\begin{abstract}

We show how to construct a large class of quantum error correcting codes, known as CSS codes, from highly entangled cluster states.  This becomes a primitive in a  protocol that \emph{foliates} a series of such cluster states into a much larger cluster state, implementing  {foliated} quantum error correction. We exemplify this construction with several familiar quantum error correction codes, and propose a generic method for decoding  foliated codes. We \tms{numerically} evaluate the error-correction performance of a family of finite-rate CSS codes known as turbo codes, finding that it performs well over \tms{moderate depth foliations}.  Foliated codes have applications for quantum repeaters and fault-tolerant measurement-based quantum computation.
\end{abstract}

\maketitle

Quantum error correction is critical to building practical  quantum information processors (QIP).   In an influential series of  papers, Raussendorf et al.\ described a  measurement based approach to fault tolerant  quantum processing using highly-entangled \emph{cluster states}, defined on a  3D lattice  \cite{raussendorf:062313,Raussendorf20062242,PhysRevLett.98.190504,1367-2630-9-6-199}.   
Raussendorf's 3D cluster state can be visualised as a \emph{foliation} of Kitaev's surface code \cite{kitaev2003ftq,dennis:4452}, i.e.\ a sequence of 2D surface code `sheets', stacked together to form a 3D  lattice.  This is evident in \cite{raussendorf:062313}, where it is shown that measuring the `bulk' qubits of a 3D cluster state leaves the two logical surface-code qubits encoded in the boundary faces in an entangled Bell-pair.

Raussendorf's 3D cluster   gained prominence for its  high fault-tolerant computational error thresholds $\lesssim1\%$.  It has applications in various QIP tasks, including long-range entanglement sharing, in which surface-code cluster states  are created at regularly spaced local nodes, which are linked by medium-range optical channels into a 3D cluster state \cite{PhysRevLett.105.250502}.  It is  capable of fault-tolerant, measurement-based quantum computation, using an elegant geometric construction that braids defects in the interior of the 3D cluster state to produce  robust Clifford gates. Universality is afforded by magic state injection and distillation \cite{bravyi_2005,PhysRevLett.98.190504,1367-2630-9-6-199}.     

The   robustness of \cite{PhysRevLett.98.190504,1367-2630-9-6-199} is inherited from  the underlying surface code, which  has a high error-correction threshold $\sim11\%$ \cite{dennis:4452,PhysRevA.81.022317,PhysRevLett.105.200502,duclos2010fast}.  The surface code has large distance,  and zero-rate  (the asymptotic ratio of the number of  logical and physical qubits), reflecting the  tradeoff between distance and rate in \tms{two} spatial dimensions 
\cite{PhysRevLett.104.050503}.
  It is  natural to ask how to adapt the foliated structure of \cite{raussendorf:062313,Raussendorf20062242} to use other underlying codes  that could \mbox{achieve a higher encoding rate.}

 Another motivation for our work is recent fault tolerant schemes that produce a universal gate set by code deformation and code switching \cite{
 bombin2014dimensional,paetznick2013universal,anderson2014fault}. Extending code foliation to  codes that circumvent magic state distillation \tms{\cite{bravyi_2005}} may produce  cluster states \tms{with lower resource overhead} for fault-tolerant measurement-based QIP.

In this letter we show that all Calderbank-Steane-Shor (CSS) codes can be \emph{clusterized}, meaning that they can be derived, \tms{using single-qubit measurements, from a larger cluster state \cite{raussendorf_2001,raussendorf_2003} defined over the code qubits plus additional ancilla qubits}.  We  use this fact to develop our main result:  generalising  Raussendorf's 3D lattice to  a foliation of {any} clusterized CSS  code. This is a larger cluster state comprised of alternating copies of a clusterized CSS code and its dual.  We demonstrate this construction for some familiar CSS codes, and present a general decoding algorithm for foliated codes, utilising the underlying code's decoder. Finally, we apply the construction to a family of \tms{finite}-rate CSS codes called turbo codes \cite{benedetto_1996,poulin2009quantum}, and present Monte Carlo simulations of the error-correction performance of foliated turbo codes.  
\emph{Background:} CSS code stabiliser generators  are classified into  two sets: 
 $S_Z \in \{I,Z\} ^{\otimes n}$ and $S_X \in \{I,X\} ^{\otimes n}$,  where $Z$ and $X$ denote the  Pauli matrices \cite{nie00}. 
 An $[[n,k,d]]$ CSS code satisfies $k=n-(|\mathcal{S}_X|+|\mathcal{S}_Z|)$.  We write a stabiliser in $\mathcal{S}_Z$ as  $Z_{\vec{b}}\equiv\otimes_{\vec{b}} Z^{b_j}$ for some binary vector $\vec{b}=(b_1,b_2,...,b_n)$ with $b_j=1$ if qubit $j$ is in the stabilizer   $Z_{\vec{b}}$, and $b_j=0$ otherwise, i.e.\ $\vec{b}$ is a row of the code's parity check matrix, $B_Z$.  Similarly, a stabilizer in $\mathcal{S}_X$ is given by  $X_{\vec{c}}$ for some binary list $\vec{c}$.  The associated {dual}  code is derived from the primal code  by exchanging  $X$ and $Z$ operators in the stabilizer generators, $X\leftrightarrow Z$.

A \emph{cluster state} is defined on a collection of qubits located at the vertices of a graph \cite{raussendorf_2001,raussendorf_2003,briegel_2009}. A qubit at vertex $v$ is associated with a cluster stabiliser $C_v=X_v(\otimes_{\mathcal{N}_v}Z)\equiv X_v Z_{\mathcal{N}_v}$, acting on it and  its neighbours,  ${\mathcal{N}_v}$.  The cluster state is the $+1$ eigenstate of the $C_v$'s.

\emph{Clusterized CSS codes:}   An $[[n,k,d]]$ CSS code can be generated from  a larger \emph{progenitor} cluster  state, i.e. \emph{clusterized}.   
 The progenitor cluster is simply the cluster state associated to the Tanner graph of $S_Z$ \tms{\cite{tanner81}}, i.e.\ a bipartite graph $G=(V,E)$ whose vertices $V$ are labeled by code qubits $j$, or ancilla qubits $a$, each associated to a stabilizer  $Z_{\mathcal{N}_{a}}\in\mathcal{S}_Z$, so that $|V|=n+|\mathcal{S}_Z|$. $E$ contains the graph edge $(a,j)$  if $[B_Z]_{a,j} = 1$.  
     We now show that a codestate of the CSS code is obtained by measuring the  ancilla qubits of the progenitor cluster  in the $X$ basis.

 In the above definition, the cluster stabiliser associated to ancilla $a$ is \mbox{$C_{a}=X_{a}Z_{\mathcal{N}_{a}}$}.  Measurement of $a$ in the $X$ basis with  outcome $\pm1$  projects  adjacent code qubits into an eigenstate of  the  code stabiliser \mbox{$ Z_{\mathcal{N}_{a}}\in\mathcal{S}_Z$}. Thus, $\mathcal{S}_Z$ is generated by  ancilla measurements.

\begin{figure}[t]
\begin{center}
\includegraphics[width=\columnwidth]{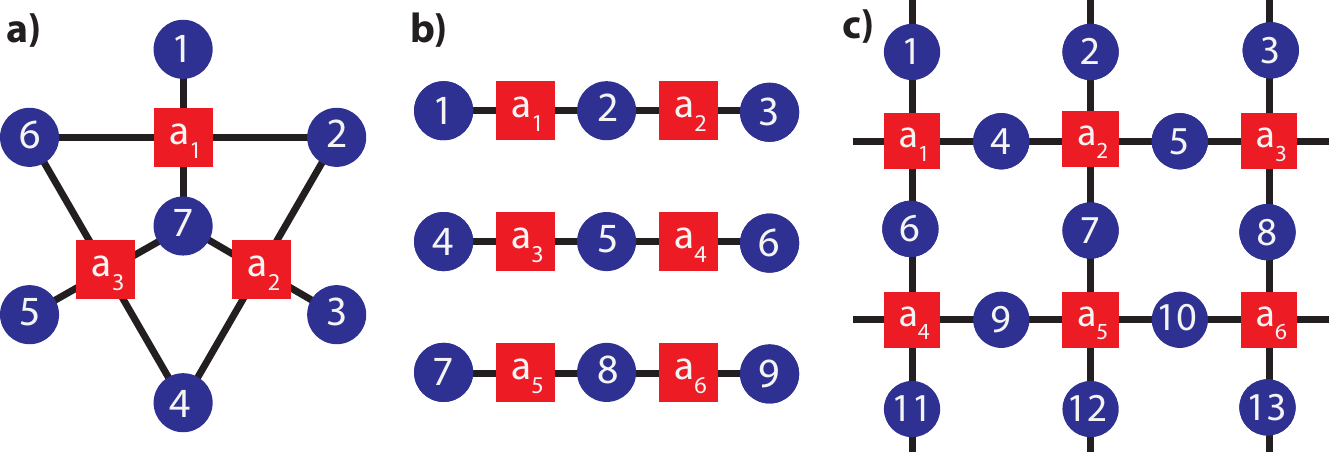}
\caption{Examples of progenitor clusters for clusterized CSS codes. {\bf a)} Clusterized Steane code. {\bf b)} Clusterized Shor code. 
 {\bf c)}  Clusterized surface code.  Code qubits (blue circles) are connected by cluster bonds (black lines) to ancilla qubits (red squares).  An $X$-basis measurement of  ancilla $a_k$  projects  neighbouring code qubits onto an eigenstate of \mbox{$\otimes_{\mathcal{N}_{a_k}} Z\in\mathcal{S}_Z$}.  } \label{codeexamples}
\end{center}
\end{figure}

Because $\mathcal{S}_X$  and $\mathcal{S}_Z$ mutually commute, the progenitor cluster  is also an eigenstate of the generators in $\mathcal{S}_X$. 
 To see this, take an element $X_{\vec{c}}\in\mathcal{S}_X$, and consider the product of cluster stabilizers centred at each code qubit $c_j\in\vec{c}$, given by $C_{\vec{c}}\equiv\otimes_{\vec{c}}C_{c_j}=\otimes_{\vec{c}}( X_{c_j}Z_{\mathcal{N}_{c_j}})$.  The neighbourhood, $\mathcal{N}_{c_j}$, of code qubit $c_j$ consists only of ancilla.  The code stabiliser $X_{\vec{c}}=\otimes_{\vec{c}} X_{c_j}$ has even overlap with any $Z$-like stabiliser, $Z_{\vec{b}_a}$ (which is generated by measurement  of ancilla $a$ in the $X$-basis).  It follows that the intersection of  $\vec{c}$ and $\vec{b}_a$ has an even number of qubits. Any ancilla qubit $a$ thus appears in the product $\otimes_{\vec{c}}C_{c_j}$ an even number of times,  so $\otimes_{a\in{\mathcal{N}_{c_j}}}Z_a=\otimes_{a\in{\mathcal{N}_{c_j}}}\mathbb{I}_a$, and  $C_{\vec{c}}= X_{\vec{c}}$. Thus, $\mathcal{S}_X$ is generated by  \mbox{cluster \tms{stabilizers}}. 

The same argument implies that logical $X$ operators of the CSS code (which are products of local $X$ operators that commute with $\mathcal{S}_Z$), are also generated by cluster stabilizers.  It follows that ancilla measurements project the cluster state into a logical $X$ codestate.

The surface code \cite{kitaev2003ftq} exemplifies the relationship between a CSS code and a progenitor cluster state. Starting from the  cluster state defined on the lattice shown in \fig{codeexamples}c, and measuring the ancilla qubits (red squares) in the  $X$ basis results in a new state on the remaining code qubits (blue circles) which is stabilised by the surface-code plaquette operators, e.g.\ \mbox{$Z_2Z_4Z_5Z_7\in\mathcal{S}_Z^\textrm{surf}$}, and vertex operators, e.g.\ $X_4X_6X_7X_9\in\mathcal{S}_X^\textrm{surf}$. It  is therefore a codestate of the surface code \footnote{We allow  $\pm1$  eigenstates of $Z_{\vec{b}_a}\in\mathcal{S}_Z$  in this definition; a trivial syndrome corresponds to agreement between $Z_{\vec{b}_a}$ and the associated ancilla $X_a$ measurement outcome.}.

Other examples of clusterized CSS codes are shown in \fig{codeexamples}a for Steane's 7-qubit  code \cite{steane_1996} 
  for which 
 \mbox{$\mathcal{S}_Z^\textrm{Steane}=\{Z_1 Z_2 Z_6 Z_7, Z_2 Z_3 Z_4 Z_7, Z_4 Z_5 Z_6 Z_7 \}$} (which is also a minimal example of the colour code \cite{duclos_2011,karigan_2010}); and  \fig{codeexamples}b for  Shor's 9-qubit code \cite{calderbank_1996,shor_1995}   for which 
  \mbox{$\mathcal{S}_Z^\textrm{Shor}\hspace{-1mm}=\hspace{-0.5mm}\{Z_1 Z_2,Z_2 Z_3, Z_4 Z_5, Z_5 Z_6, Z_7 Z_8, Z_8 Z_9 \}$}.

The examples in \fig{codeexamples} illustrate the fact that $X$-measurements of the ancilla qubits project out code stabilizers in $\mathcal{S}_Z$, while each stabiliser in $\mathcal{S}_X$ comes `for free' simply by considering products of $C_j$'s acting on the corresponding code qubits, and noting that these products act trivially on the ancilla qubits.  For example, in the Steane code cluster of \fig{codeexamples}a, it is straightforward to check that  \mbox{$C_1C_2C_6C_7=X_1X_2X_6X_7\in\mathcal{S}_X^\textrm{Steane}$}. 

\begin{figure}[t]
\begin{center}
\includegraphics[trim=0cm 0cm 0cm 0cm, clip=true,width=\columnwidth]{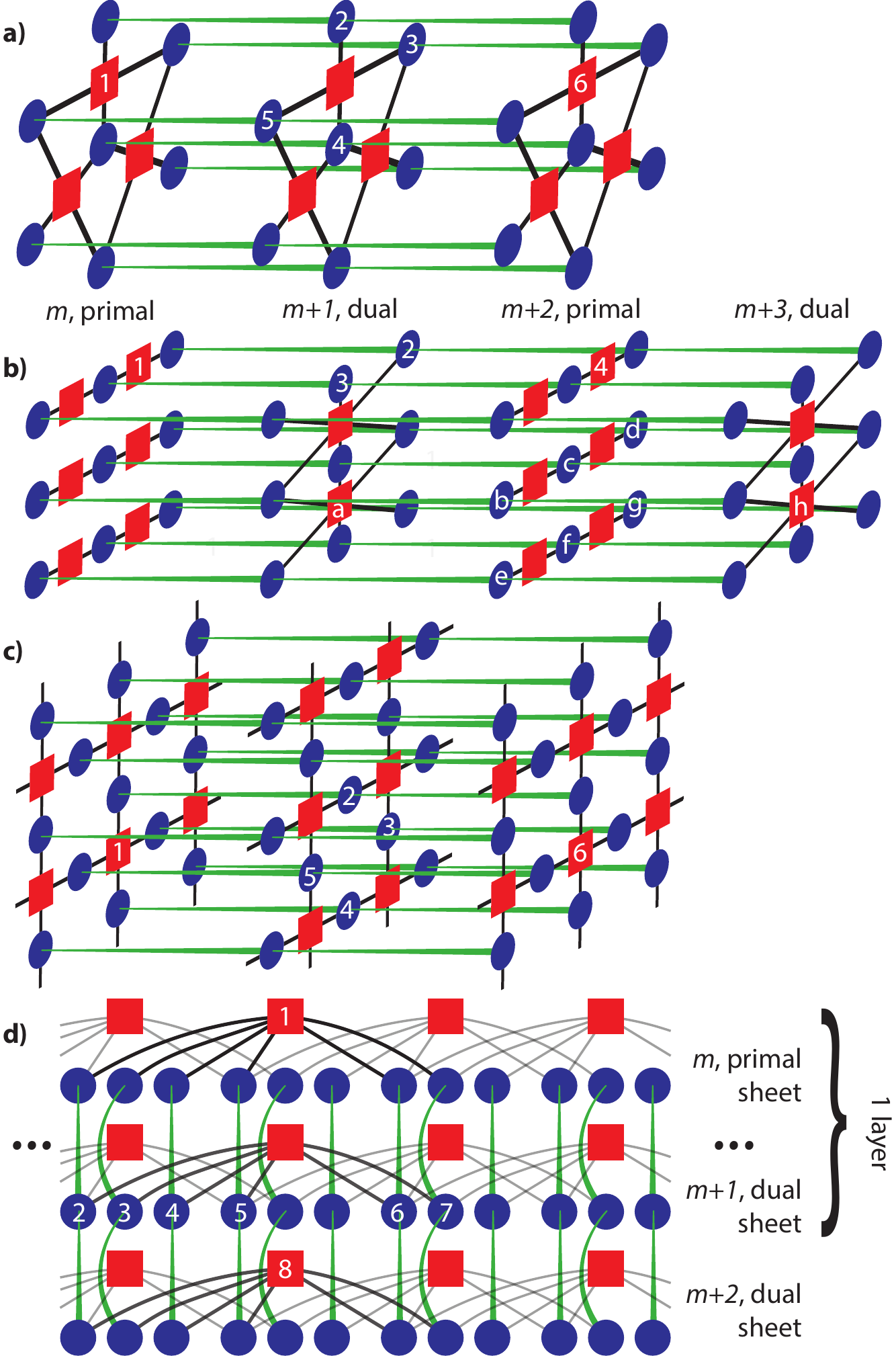}
\caption{Examples of foliated, clusterized CSS codes. Code qubits (blue circles) share cluster bonds (black lines) with  ancilla qubits (red squares) in the same  sheet, and with code qubits in  adjacent sheets (green lines). 
{\bf a)} Foliated Steane code.  Being self-dual, primal and dual sheets are identical.  The product of cluster stabilizers centred on the numbered qubits generates parity check operators \mbox{$C_1 C_2...C_6= X_1 X_2...X_6$}. {\bf b)} Foliated Shor code.  This code is not self-dual, so   primal and dual sheets are different, and there are two kinds of parity check operators, \mbox{$C_aC_b...C_h= X_aX_b...X_h$} centred on  primal sheets, and $C_1C_2C_3C_4= X_1X_2X_3X_4$ centred on dual sheets. {\bf c)} Foliated surface code \cite{raussendorf:062313}, with parity check operators \mbox{$C_1 ...C_6= X_1 ...X_6$}. 
{\bf d)} Foliated self-dual convolutional code with parity check operator \mbox{$C_1...C_8= X_1...X_8$}. 
Stabilizers are generated by translations of the kernel (indicated by  thick  edges) across frames (here, the frame length is 3).
}
 \label{foliatedexample}
  \label{convex}
\end{center}
\end{figure}

\emph{Foliated codes:} Raussendorf's 3D cluster state construction \cite{Raussendorf20062242,PhysRevLett.98.190504,raussendorf:062313,1367-2630-9-6-199}, \fig{foliatedexample}c, can be viewed as a \emph{foliation} of the surface code cluster state shown in \fig{codeexamples}c.  Alternating `sheets' of the primal surface code cluster state and its dual  are stacked together \footnote{The dual to the surface code is also a surface code, albeit on the lattice geometrically rotated by 90 degrees.}, with additional cluster bonds (green lines) extending between  code qubits in each sheet and the corresponding  code qubits in the adjacent dual  sheets.

We now generalise this construction to arbitrary CSS codes. Take an alternating stack of sheets of clusterized primal and dual codes,  
  and link the  sheets together by creating additional cluster bonds between primal code qubits in a given sheet, $m$, and the corresponding dual code qubits in the adjacent sheets, $m\pm1$.  We call this a \emph{foliated  code}.  The number of \emph{layers}, $L$, in the foliated construction counts the number of primal--dual sheet pairs, so that $1\leq m\leq 2L+1$.
  
  \fig{foliatedexample}a shows the example of a foliated  Steane  code, (which is self-dual, so the primal and dual sheets are  identical). 
  \fig{foliatedexample}b shows the foliated Shor  code, for which primal and dual clusters  are different. 
One can readily verify that this definition preserves the key feature of Rausendorf's construction: measuring the bulk qubits and the boundary ancilla qubits leaves the two boundary sheets in an encoded Bell-state \cite{raussendorf:062313}. Because this feature enables fault-tolerant measurement-based quantum computation and long-range entanglement sharing, our generalization has immediate applications in these settings, offering additional flexibility in the choice of code.

\emph{Errors:} Errors may arise during construction of the cluster,  storage of the qubits, or during single-qubit measurement.   
 In the error models we consider, preparation and measurement errors can be mapped to possibly correlated storage errors \cite{PhysRevA.81.022317}.   Furthermore, after the cluster is created, since we perform  single qubit $X$ measurements,  $X$ errors do not affect the measurement outcome; only  $Z$ errors on the final foliated cluster act nontrivially.  We note that  $X$ errors during cluster construction are equivalent to correlated $Z$ errors in the final cluster \cite{PhysRevLett.105.200502,PhysRevA.81.022317}.  This asymmetry between $X$ and $Z$ errors is a consequence of the  asymmetry in the definition of the cluster stabilizers.  \tms{Correlated or asymmetric errors may also arise in specific applications, such as long-range repeaters where inter-node quantum transmission errors are much worse than those within a node, which can be mitigated by suitable choice of code  \cite{cafaro}.}

\emph{Parity check  operators:} Errors in the foliated cluster are detected by parity check operators: a $Z$ error will flip one or more parity checks, giving a non-trivial error syndrome for the foliated cluster.   Importantly, the parity check measurement outcomes can be inferred from sets of single-qubit $X$ measurements.   

Each parity check operator is associated with a CSS code stabiliser within a code sheet.  To  construct a parity check operator, consider the CSS code stabiliser $X_{\vec{c},m}\in\mathcal{S}_X$, in  sheet $m$ of a foliated cluster state.  The product of foliated cluster stabilizers centred on each of the code qubits indicated by $\vec{c}$ is \mbox{$C_{\vec{c},m}= X_{\vec{c},m}Z_{\mathcal{N}_{{\vec{c},m}}}= Z_{\vec{c},{m-1}}X_{\vec{c},m}Z_{\vec{c},{m+1}}$}. 
 The dual code sheets, $m\pm1$, each have a cluster stabilizer \mbox{$C_{a_{\vec{c}},{m\pm1}}=X_{a_{\vec{c}},{m\pm1}}Z_{\vec{c},{m\pm1}}$} centred on an ancilla qubit $a_{\vec{c}}$ associated to ${\vec{c}}$.  Thus,  {$\hat P_{\vec{c},m}\equiv C_{a_{\vec{c}},{m-1}}C_{\vec{c},m}C_{a_{\vec{c}},{m+1}}= X_{a_{\vec{c}},{m-1}}X_{\vec{c},m}X_{a_{\vec{c}},{m+1}}$} defines a parity check for the foliated cluster,  centred on  code stabilizer $X_{\vec{c},m}$.  Note that parity check operators centred on primal sheets  share no common qubits with those centred on dual sheets. 
      
This generalises the construction of the parity check operators for  Raussendorf's 3D cubic lattice, which are formed by products of $X$ operators on the faces of the cubic unit cells, as shown in \fig{foliatedexample}c (exemplified by numbered qubits).  Parity check operators for other foliated CSS codes are exemplified by labelled qubits in other panels of \fig{foliatedexample}. 
 In a non-self-dual code, such as the Shor code,  primal and dual parity check operators  may have different weights, \fig{foliatedexample}b.

Logical code operators within a sheet commute with the parity check operators.  It follows that for an underlying $[[n,k,d]]$ code, there are weight-$d$ undetected error chains on the foliated cluster, as in \cite{Raussendorf20062242,PhysRevLett.98.190504,raussendorf:062313,1367-2630-9-6-199,dennis:4452}.  Since the structure of the code in the direction of foliation is a simple repetition, it follows that the foliated cluster inherits the distance of \mbox{the underlying code.}

 \emph{Decoding:}  A non-trivial error syndrome indicates the presence of $Z$ errors.  
 If the error probability is sufficiently small, the most likely class of errors can be  inferred from the syndrome with high probability, facilitating error recovery.  Small codes can be decoded by brute-force, but this is not computationally scalable in $n$. 
 
 There are a number of computationally efficient, near-optimal   decoders available for both the 2D surface code and its 3D foliation, including hard decoders (which return a specific high-likelihood error pattern) based on  perfect matching \cite{dennis:4452,PhysRevA.81.022317}, and soft decoders (which return a probability distribution over error patterns)  based on renormalisation methods \cite{duclos2010fast,duclos2014fault}.  
 
Surface code decoders naturally generalise to the 3D Raussendorf lattice, as exemplified by matching-based decoders.  
While generic CSS codes cannot typically be efficiently decoded, many exact or heuristic decoders are known for specific code constructions  \cite{dennis:4452,PhysRevLett.91.177902,poulin2009quantum,duclos2010fast}. The problem we address here is  to use 
 a soft decoder for the underlying CSS  code -- which we presume to be efficient -- as a subroutine in a decoder for the foliated construction.   We describe  a heuristic method based on belief propagation  (BP) that may work in many cases \cite{mceliece_1998,poulin2008iterative}.  
 We assume the existence of soft decoders for the underlying CSS primal  and dual codes, which, given a physical error model,  calculates the  probability of a Pauli error $\sigma$ on code qubit $j$, $P(\sigma_j | S_{\textrm{CSS}})$, conditioned on  a syndrome, $S_{\textrm{CSS}}$, which may itself be unreliable.

  In the foliated case, consider a parity check operator $\hat P_{\vec{c},m}=X_{a_{\vec{c}},m-1}X_{\vec{c},m}X_{a_{\vec{c}},m+1}$, centred on primal sheet $m$. A non-trivial syndrome can arise because of errors on  code qubits $\vec{c}$  within  code sheet $m$, or due to errors on the corresponding ancilla qubits, $a_{\vec{c}}$ in adjacent dual sheets $m\pm1$.  If the dual-sheet ancilla qubits were error-free, then all the parity check failures would be due solely to in-code qubit errors, so that the parity check outcomes centred on sheet $m$ would be in direct correspondence with the CSS code syndrome for that sheet.  The code syndrome could then be used in the CSS  decoder to calculate a soft  error model on sheet $m$, from which an error correction strategy could be determined.  
  
 However, errors on the dual-sheet ancilla qubits mean that the in-sheet syndrome passed to the CSS decoder is itself unreliable.  To account for dual-sheet ancilla errors, we embed the CSS decoder in a BP routine, as follows. 
 \newline
\textbf{Step 1}:  For each code qubit, $j$, in sheet $m$, the CSS decoder calculates an in-sheet error model probability distribution, \mbox{$P_m(\sigma_j|S_{m}\cup P_{m\pm1}(a_k))$},  subject to both the measured code syndrome, $S_{m}$, which is derived from the foliated parity-check operators centred on sheet $m$, and an assumed error model, $P_{m\pm1}(a_k)$, for errors on  ancilla qubits, $a_k$, in  adjacent dual sheets.  
 \mbox{\textbf{Step 2}: Using the result of Step 1 we  fix the code qubit} error model, $P_m(\sigma_j)$, and calculate an error model on the dual-sheet ancilla qubits, $P_{m\pm1}(a_k|S_{m\pm1}\cup P_{m}(\sigma_j))$. 
 \mbox{\textbf{Step 3}: We iterate  Step 1}, using the result of Step 2 for $P_{m\pm1}(a_k)$,  repeating 
 until each error model converges.

 \begin{figure}[t]
\begin{center}
\includegraphics[width=\columnwidth]{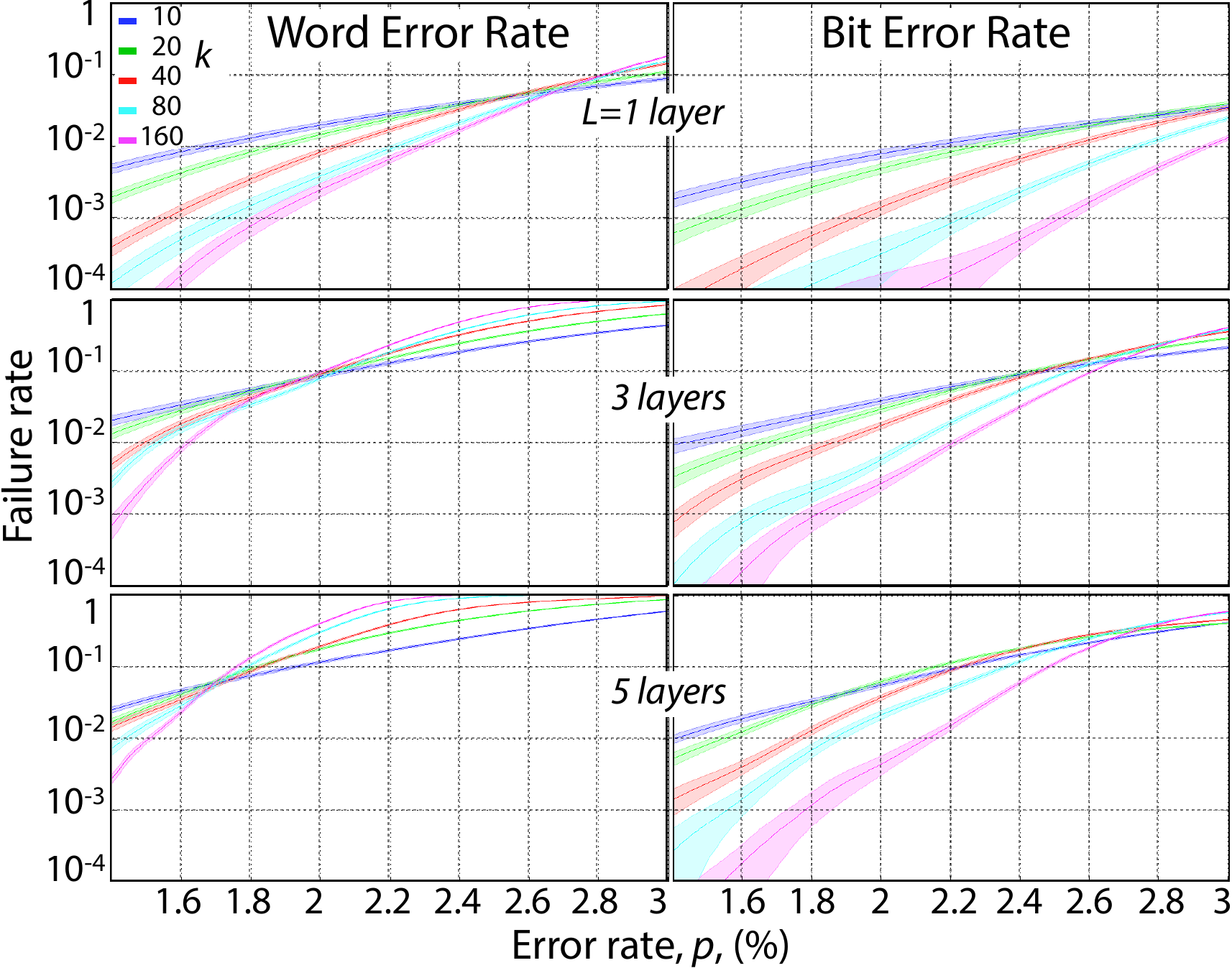}
\caption{Numerical performance results for a foliated $r=\frac{1}{25}$, $d=25$ turbo code, for different numbers of foliated layers, $L$ (rows).  Different colours correspond to different code sizes, $k=n r$; shading indicates $\pm1\sigma$.  A `layer' consists of a primal code sheet and a dual code sheet (see \fig{convex}d).  Word Error Rate (left column) counts any error(s) across all $k$ logical qubits.  Bit Error Rate (right column) counts the failure rate per logical qubit.} 
\label{turbonumerics}
\end{center}
\end{figure}

\emph{Turbo  Codes:}  We now consider the  
class of  turbo  codes, which are finite-rate CSS codes with bounded-weight stabilizers \cite{poulin2009quantum,mceliece_1998,papaharalabos_2007}.  These are capable of encoding an arbitrary  number of logical qubits with finite rate, $r=k/n$.  Essentially, turbo codes are formed from a concatenation of two \emph{convolutional} codes: an inner code $G_I$, and an outer code $G_O$ \cite{PhysRevLett.91.177902,ollivier_2009,tillich_2009,tan2010efficient}, 
each of which can be decoded  with soft trellis decoders 
\cite{mceliece_1998,papaharalabos_2007,poulin2009quantum,tan2010efficient,geldmacher_2012}.   

Convolutional codes are defined over an ordered set of qubits. The  code stabilizers are generated by a kernel which is repeatedly translated across \emph{frames} (i.e.\ blocks of the underlying physical qubits).   
 For illustrative purposes, \fig{convex}d shows a foliation of three sheets of a $d=3$, $r=1/3$, weight-6 self-dual CSS convolutional code cluster.  The code stabiliser kernel is indicated by the dark cluster edges within a sheet.  Turbo codes are conceptually similar, albeit with more complicated Tanner graphs.

Turbo codes provide a platform for testing the foliated construction on codes that are quite different to the surface code.  Since they are a code family, we  analyse the performance of the codes as a function of the code size $k=nr$, and the number of foliated layers, $L$. A soft trellis decoder \cite{tu_2005,tan2010efficient} for the underlying code is embedded as a subroutine in a BP decoder spanning the sheets of the foliation.    The BP decoder run-time is linear in $L$, however the trellis decoder complexity is exponential in the size of the turbo code frame length, making  simulations practically slow.

\fig{turbonumerics} shows the performance of a $d=25$,  $r={1}/{25}$,  self-dual foliated turbo code, based on Monte Carlo  simulations of errors.  \tms{As noted earlier, $X$ errors on the foliated cluster  commute with parity check measurements.  Thus, for our simulations we assume a phenomenological error model in which uncorrelated $Z$ errors are distributed independently across the  cluster with probability $p$.}   The decoder performance is quantified in terms of both word error rate (WER), which is the probability of one or more errors across all $k$ encoded qubits, and the bit error rate (BER) which is the probability of an error in each of the encoded qubits.  

For each $L$, there is a threshold error rate around $p\sim 2\%$, below which the code performance improves with code length (up to  at least 160 encoded logical qubits per code sheet), consistent with pseudo-threshold behaviour seen in turbo codes \cite{poulin2009quantum}.    As $L$ increases, the threshold decreases, more pronouncedly for the WER than the BER.  The range of $k$ and $L$ that we can simulate is limited by computational time, so we cannot explore the asymptotic performance  for   large $L$.  Nevertheless, numerics indicate that foliated turbo codes perform quite well for moderate depth foliations.

 We note  that the foliated construction  transforms a clusterised code into a fault tolerant resource state, but with a consequent reduction in threshold.  This  is seen in \fig{turbonumerics}, and in Raussendorf's construction in which the fault-tolerant threshold $\lesssim1\%$ is smaller than the $\sim11\%$ threshold for the surface code  on which it is based.  The $\sim1\%$ threshold observed for foliated surface codes is obtained by scaling the code distance $d$ and foliation depth $L$ together. Here, the code distance is fixed at \mbox{$d=25$,} which is responsible for the observed decreasing value of the pseudo-threshold with increasing $L$.

Our main motivation for studying turbo codes is  to demonstrate the foliated construction and BP decoder in an extensible, finite-rate code family.  Practically, these and other finite-rate codes may have  applications in \tms{fault-tolerant  quantum repeaters networks  \cite{PhysRevLett.105.250502,bib:Duan01},  where  local nodes  create optimal clusterized  codes 
to reduce resource overheads or error tolerance \cite{satoh}}.  

In conclusion, we have shown how to clusterize arbitrary CSS codes.  We have shown how to foliate  clusterized codes, generalising Raussendorf's 3D foliation of the surface code.  We have described a generic approach to decoding errors that arise on the foliated cluster using an underlying soft decoder for the  CSS code as a subroutine in a BP decoder, and applied it  to  error correction \tms{by means} of a  foliated turbo code.  This construction may have  applications where codes with finite rate are useful, such as long-range quantum repeater networks.

\acknowledgments{
This work was  funded by the ARC Centre of Excellence for Engineered Quantum Systems, NSERC and the Canadian Institute for Advanced Research.  We thank Sean Barrett, Andrew Doherty, Terry Rudolph, Clemens Mueller and Stephen Bartlett for helpful discussions.
}

\bibliography{bib,bibliography}

\end{document}